\documentclass[12pt,twoside]{article}
\usepackage{fleqn,espcrc1,epsfig}

\title{Thermal and chemical equilibration of hadronic matter
\thanks{Work supported by BMBF, GSI Darmstadt and DFG.}}

\author{E. L. Bratkovskaya, W. Cassing, C. Greiner,
	 M. Effenberger, U. Mosel and A. Sibirtsev\\[3mm]
Institut f\"{u}r Theoretische Physik, Universit\"{a}t Giessen,
	  35392 Giessen, Germany}
\begin{document}

\maketitle

\begin{abstract}
We study thermal and chemical equilibration in 'infinite' hadron matter
as well as in finite size relativistic nucleus-nucleus collisions using
a BUU cascade transport model with resonance and string
degrees-of-freedom.  The 'infinite' hadron matter is simulated within a
cubic box employing periodic boundary conditions.  The various equilibration
times depend on baryon density and energy density and are much shorter
for particles consisting of light quarks then for particles including
strangeness.  For kaons and antikaons the chemical equilibration time
is found to be larger than $\simeq$ 40 fm/c for all baryon and energy
densities considered.  The inclusion of continuum excitations, i.e.
hadron 'strings', leads to a limiting temperature of $T_s\simeq$ 150 MeV.
\end{abstract}

\section{INTRODUCTION}

Nucleus-nucleus collisions at relativistic and ultrarelativistic
energies are studied experimentally and theoretically to obtain
information about the properties of hadrons at high density and/or
temperature as well as  about the phase transition to a new state of
matter, the quark-gluon plasma (QGP).  In the latter deconfined partons
are the essential degrees of freedom that resolve the underlying
structure of hadrons \cite{QM}.  Whereas the early 'big-bang' of the
universe most likely evolved through steps of kinetic and chemical
equilibrium, the laboratory 'tiny bangs' proceed through phase-space
configurations that initially are far from an equilibrium phase and
then evolve by fast expansion. These 'specific initial conditions' --
on the theoretical side -- have lead to a rapid development of
nonequilibrium quantum field theory and nonequilibribrium kinetic
theory \cite{BotMal90,Henning}.  Presently, semiclassical transport
models are widely used as approximate solutions to these theories and
practically are an essential ingredient in the experimental data
analysis. For recent reviews we refer the reader to Refs.
\cite{Ko,Bass,Cass99}.

On the other hand, many observables from strongly interacting systems
are dominated by many-body phase space such that spectra and abundances
look 'thermal'.  It is thus tempting to characterize the experimental
observables by global thermodynamical quantities like 'temperature',
chemical potentials or entropy \cite{BM,Satz,Sollfrank,Spieles,Cleymans}.
We note, that even the use of macroscopic models like hydrodynamics
\cite{Hydro,Rischke} employs as basic assumption the concept of local
thermal and chemical equilibrium. The crucial question, however, how
and on what timescales a global thermodynamic equilibrium can be
achieved, is presently a matter of debate. Thus nonequilibrium
approaches have been used in the past to address the problem of
timescales associated to global or local equilibration
\cite{Rafelski,Cass90,Lang91,Bl93,Brav1,Brav2,Brav3,Solfr99}.  In view
of the increasing 'popularity' of thermodynamic analyses a thorough
microscopic reanalysis of this questions appears necessary especially
for nucleus-nucleus collisions at ultrarelativistic energies that aim
at 'detecting' a phase transition to the QGP.

In this contribution we discuss equilibration phenomena in 'infinite'
hadronic matter using a microscopic transport model that contains both
hadron resonance and string degrees-of-freedom.  With this
investigation we want to provide insight into the reaction dynamics by
the use of cascade-like models and also point out some of their
limitations.  The 'infinite' hadronic matter is modelled by
initializing the system solely by nucleonic degrees of freedom through
a fixed baryon density and energy density, while confining it to a
cubic box and imposing periodic boundary conditions during the
propagation in time.

\section{EQUILIBRATION AND LIMITING TEMPERATURE}

To investigate the equilibration phenomena addressed above we perform
microscopic calculations using the Boltzmann-Uehling-Uhlenbeck (BUU)
model of Refs. \cite{Effe99gam,EffePhD}. This model is based on the
resonance concept of nucleon-nucleon and meson-nucleon interactions at
low invariant energy $\sqrt{s} \ $ \cite{TeisZP97}, adopting all
resonance parameters from the Manley analysis \cite{Manley}.  The high
energy collisions -- above $\sqrt{s}$ = 2.6~GeV for baryon-baryon
collisions and $\sqrt{s}$ = 2.2~GeV for meson-baryon collisions -- are
described by  the LUND string fragmentation model FRITIOF
\cite{FRITIOF}. This aspect is similar to that used in the HSD approach
\cite{Cass99,Ehehalt,Brat98,Geiss} and the UrQMD code \cite{Bass}. For
a detailed description of the underlying model at low energy we refer
the reader to Ref.~\cite{EffePhD}.

\subsection{A box with periodic boundary conditions}

In order to study 'infinite' hadronic matter problems we confine the
particles in a cubic box with periodic boundary conditions for their
propagation similar to a recent box calculation within the UrQMD model
\cite{Brav1}. We specify the initial conditions, i.e. baryon density
$\rho$, strange particle density $\rho_S$ and energy density
$\varepsilon$ as follows: first the initial system is fixed to $N_p=80$
protons and $N_n=80$ neutrons, which are randomly distributed in a
cubic box of volume $V$. The 3-momenta $\vec p_i$ of the nucleons in a
first step are randomly distributed inside a Fermi-sphere of radius
$p_F$ = 0.26~GeV/c (at $\rho_0$) and in a second step  boosted by $\pm
\beta_{cm}$ by a proper Lorentz transformation. Thus the initial baryon
density $\rho$ is fixed as $\rho=A/V$, $A=N_p+N_n$. The strange
particle density is set to zero as in related heavy-ion experiments
while the energy density is defined as $\varepsilon = E/V$, where $E$
is the total energy of all nucleons
\begin{eqnarray}
E=\sum\limits_i^A\sqrt{p_i^2+m_N^2}.
\label{energy_i}
\end{eqnarray}
The  boost velocity $\beta_{cm}$ is related to the initial energy density
$\varepsilon$ (excluding Fermi motion) as
\begin{eqnarray}
\beta_{cm}=\sqrt{1-{\rho^2 m_N^2\over \varepsilon^2} }
\label{betta}\end{eqnarray}
using
$\varepsilon = \gamma_{cm} \rho m_N$
with $\gamma_{cm}= 1/\sqrt{1-\beta_{cm}^2}$. Recall that $\rho_0 m_N \simeq
0.15$~GeV/fm$^3$ so that an energy density $\varepsilon \simeq
1.5$~GeV/fm$^3$ at density $\rho_0$ corresponds to $\gamma_{cm}\simeq
10$, i.e. the SPS energy $T_{lab}\simeq 185$~A$\cdot$GeV. We thus start
with a 'true' nonequilibrium situation in order to mimique the  initial
stage in a relativistic heavy-ion collision. The initial phase
represents two interpenetrating, (ideally) infinitely extended fluids
of cold nuclear matter.

We now propagate all particles in the box in the cascade mode (without
mean-field potentials) using periodic boundary conditions, i.e.
particles moving out of the box are reinserted at the opposite side
with the same momentum. The phase-space distribution of particles then
can change due to elastic collisions,  resonance and string production
and their decays to mesons and baryons again. We recall that we include
all baryon resonances up to an invariant mass of 2 GeV and meson
resonances up to the $\phi$-meson. According to the initial conditions
for $\varepsilon$ and $\rho$ the factor $\gamma_{cm}$
determines if strings are excited in the very first collisions. This is
the case for $\gamma_{cm} > 1.4$ where the early equlibration stages
are dominated by string formation and decay.

\subsection{Chemical equilibration}

\begin{figure}[h]
\centerline{\psfig{figure=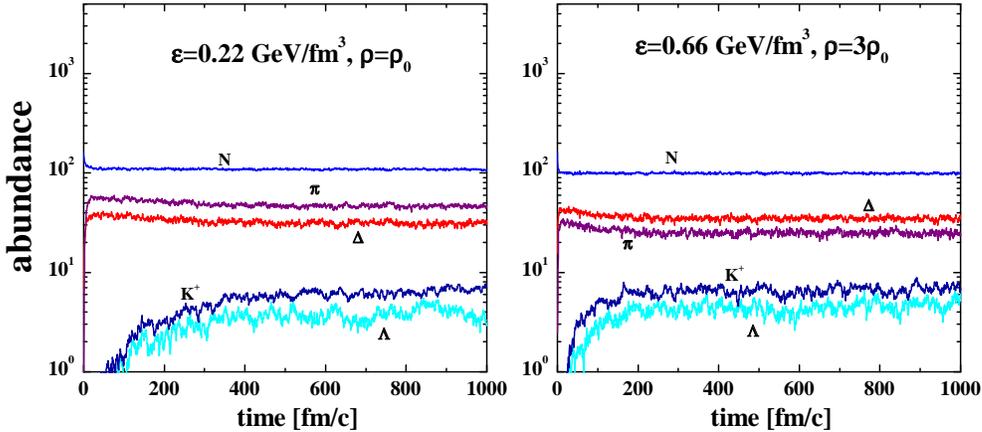,width=13cm}}
\vspace*{-5mm}
\caption{Time evolution of the various
particle abundances (nucleons $N$,  $\Delta$,  $\Lambda$, $\pi$
and  $K^+$ mesons) for density $\rho=\rho_0$ (left
panel) at energy density $\varepsilon =0.22$ GeV/fm$^3$ and for
$\rho=3\rho_0$ (right panel) at $\varepsilon =0.66$ GeV/fm$^3$.}
\label{Fig1}
\end{figure}

Figure \ref{Fig1} shows the time evolution of the various particle
abundances (nucleons $N$,  $\Delta$,  $\Lambda$, $\pi$ and  $K^+$
mesons) for density $\rho=\rho_0$ (left panel) at
energy density $\varepsilon =0.22$ GeV/fm$^3$ and for
$\rho=3\rho_0$ (right panel) at $\varepsilon =0.66$~GeV/fm$^3$.
These initial conditions correspond to bombarding
energy $T_{lab}$ per nucleon of roughly 2 A$\cdot$GeV.
In Fig.~\ref{Fig1} we count all particles which are
'hadronized', i.e.\ produced by string decay after a  formation time of
$\tau_F=0.8$~fm/c in their rest frame.

After several fm/c the number of nucleons decreases due to inelastic
collisions that produce either baryon resonances or additional mesons.
The number of $\Delta$-resonances grows up to a maximum in a few fm/c,
since a lot of $\Delta$'s are produced in the first $NN$ collisions;
their number subsequently decreases with time due to their decay and
excitation of further resonances or due to reabsorption.  The numbers
of $\pi$'s  and $\eta$'s increase very fast and reach the equilibrium
value within a few fm/c whereas the strange particles ($K^+,
K^-,\Lambda$) require a much longer time for equilibration.

For the higher energies the initial particle production proceeds via
the formation and decay of string excitations. This leads in particular
to a very early onset of strange particles (mainly kaons and hyperons)
within the first fm/c either due to the initial strings or due to
secondary or ternary baryon-baryon, meson-baryon and meson-meson
induced string-like interactions.
In Ref. \cite{Geiss} it was shown that
these early secondary and ternary reactions can contribute up to about
50 $\%$ of the total strange particles obtained in a Pb~+~Pb reaction at
CERN SPS energies and thus explain the factor of 2 in the observed
relative strangeness enhancement compared to p+p reactions. This,
however, does not imply that chemical equilibrium for the dominant
strange particles has been achieved in this reaction, as our analysis
clearly shows. In the later stages, when the system has become, more or
less, isotropic in momentum space, strange particles can only be
further produced by low energy hadronic reactions, which, however, have
a considerable threshold and are thus strongly suppressed. This
explains the long chemical equilibration times for the strange
particles first demonstrated by Koch, M\"uller and Rafelski
\cite{Rafelski}.

In order to define an overall chemical equilibration time
we perform a fit to
the particle abundances $N(t)$ for pions and kaons as
\begin{eqnarray}
N(t) = N_{eq} \left(1 - \exp(-t/\tau_{eq})\right),
\label{taueq}\end{eqnarray}
where $N_{eq}$ is the equilibrium limit.  The equilibration time $\tau_{eq}$
thus corresponds to the time $t$ when $\simeq 63$\% of $N_{eq}$ is achieved.

\begin{figure}[h]
\centerline{\psfig{figure=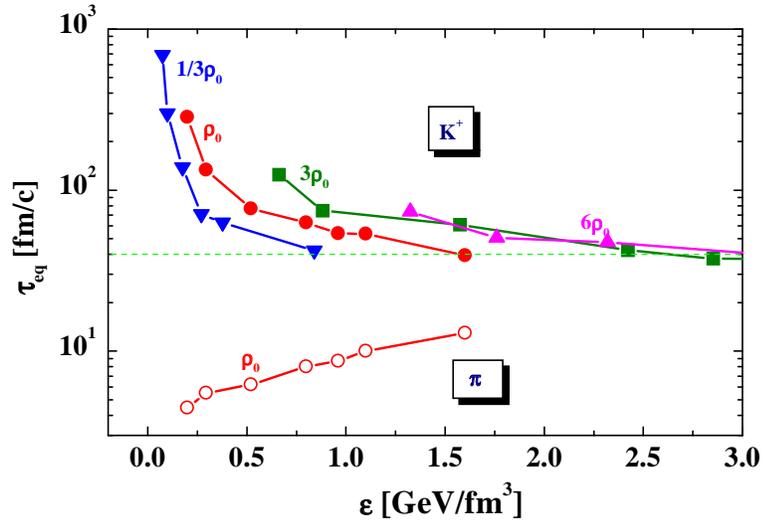,width=10cm}}
\vspace*{-5mm}
\caption{Equilibration time $\tau_{eq}$ versus energy density $\varepsilon$
for $\pi$ and $K^+$ mesons at different baryon densities $1/3\rho_0,
\rho_0, 3\rho_0$ and $6\rho_0$.}
\label{Fig4}
\end{figure}

Figure \ref{Fig4} shows the equilibration time $\tau_{eq}$ versus
energy density for $\pi$ and $K^+$ mesons at different baryon densities
of $1/3\rho_0, \rho_0, 3\rho_0$ and $6\rho_0$. We find that the
equilibration time for pions scales as $\tau_{eq}^\pi \sim 1/\rho$ or
$\Gamma_\pi \sim \rho$, thus we present the curve only for baryon
density $\rho_0$.  Whereas $\tau_{eq}^\pi$ slowly grows with
energy-density, $\tau_{eq}^K$ falls steeply with $\varepsilon$. This
marked difference is due to the fact that, on one hand, the kaon
production rate increases dramatically with $\sqrt{s}$ whereas that of
the pions, on the other hand, is more flat. With increasing energy thus
more strange particles are produced through strings especially from the
primary collisions with high $\sqrt{s}$ and the chemical equilibration
is achieved faster.

In Fig.~\ref{Fig4} we have considered an 'ideal' situation, i.e. hadron
matter at fixed energy and baryon density.  In realistic heavy-ion
collisions the system goes through the different stages due to
interactions and expansion. However, as follows from Fig. \ref{Fig4},
the equilibration time for strangeness is larger than 40~fm/c for all
energy and baryon densities. Thus in realistic nucleus-nucleus
collisions the chemical equilibration of strange particles requires
also a time above~40 fm/c which is considerably larger than the actual
reaction time of a few 10 fm/c or less \cite{ThermoNPA}.

The particle abundances used to extract $\tau_{eq}$ in Fig. \ref{Fig4}
have been calculated without any in-medium potentials.
In fact, the introduction of attractive potentials (especially for
$K^-$) will lower the hadronic thresholds and thus increase the
scattering rate between strange and nonstrange hadrons, whereas the
$K^+$ feels some repulsive potential and the trend goes in the opposite
way.  According to our calculations such in-medium modifications (in
line with Ref. \cite{Cass99}) give a correction to the $K^+$
equilibration times by atmost 10 \% and shortens the $K^-$
equilibration times up to 20 \% at density $\rho_0$.

\subsection{Thermal equilibration and limiting temperature}

In this subsection we investigate the approach to thermal equilibration.
For the equilibrated system we can extract a temperature $T$ by fitting the
particle spectra with the Boltzmann distribution
\begin{eqnarray}
{d^3N_i\over dp^3} \sim \exp(-E_i/T),
\label{Boltz}\end{eqnarray}
where $E_i=\sqrt{p_i^2+m_i^2}$ is the energy of particle $i$.  We note
that at the temperatures of interest here, the Bose and Fermi
distributions are practically identical to a Boltzmann distribution. We
find that in equilibrium  the spectra of all particles can be
characterized by one single temperature $T$ \cite{ThermoNPA}.

\begin{figure}[h]
\centerline{\psfig{figure=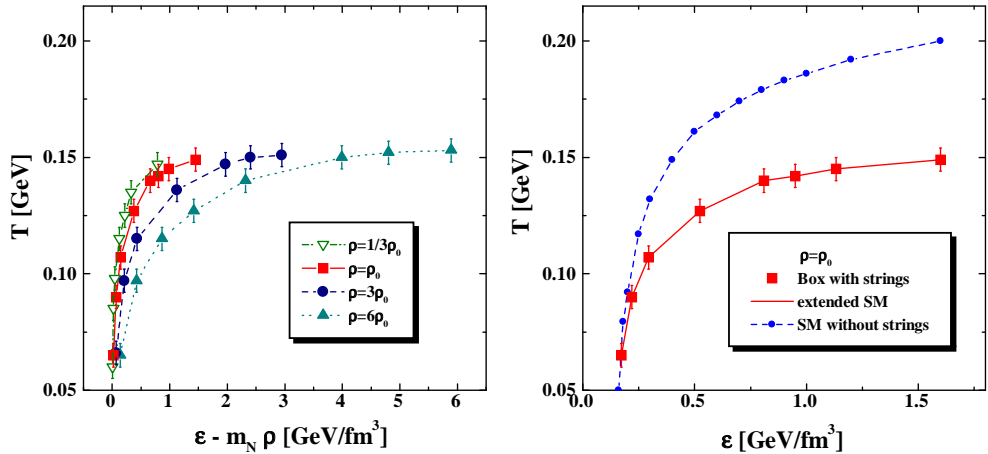,width=13cm}}
\vspace*{-5mm}
\caption{Left panel: equilibrium temperature $T$ versus the energy
density  $\varepsilon-m_N\rho$ for different baryon densities $\rho$:
$1/3\rho_0$ (open down triangles), $\rho_0$ (full squares), $3\rho_0$
(full dots), $6\rho_0$ (full up triangles).
Right panel: equilibrium temperature $T$ versus the energy density for
baryon density $\rho=\rho_0$. The full dots  correspond to the
statistical model (SM) without strings, the full squares show our
box calculations including string degrees of freedom, while the
solid line shows the result from the extended SM including a Hagedorn
mass spectrum for strings.}
\label{Fig7}
\end{figure}

In the left panel of Fig.~\ref{Fig7} we display the
temperature $T$ versus energy density $\varepsilon$
for different baryon densities $\rho$:  $1/3\rho_0$
(open down triangles), $\rho_0$ (full squares), $3\rho_0$ (full dots),
$6\rho_0$ (full up triangles). In order to compare calculations for
different baryon densities we have subtracted the baryon energy density
at rest, i.e. $\simeq m_N\rho$ (except for Fermi motion). As seen from
Fig.~\ref{Fig7} the temperature grows with energy density up to a
limiting value reminiscent of a 'Hagedorn' temperature \cite{Hagedorn}.
 From our detailed investigations we obtain for the limiting temperature
$T_s \simeq 150\pm 5$~MeV which practically does not depend on baryon
density.  Such a singular behavior of $\varepsilon(T)$ for $T\simeq
T_s$ has also been found in the box calculations in Ref.~\cite{Brav1}
for $\rho=\rho_0$.  Our limiting temperature is slightly higher than
that in Ref.~\cite{Brav1} ($T_s = 130 \pm 10$~MeV)  due to the different
number of degrees of freedom; the model \cite{Brav1} contains more
resonances and uses a different threshold for string excitations.
Thus, there is some phenomenological sensitivity to the hadronic zoo of
particles and string thresholds employed in the model.

In order to investigate the equilibrium behavior of hadron matter we
also compare our transport (box) calculations with a simple Statistical
Model (SM) for an Ideal Hadron Gas where the system is described
by a grand canonical ensemble of non-interacting fermions and bosons in
equilibrium at temperature $T$.  All baryon and meson species
considered in the transport model \cite{Effe99gam} also have been
included in the statistical model \cite{ThermoNPA}.

Within the SM we find that the temperature increases continuously with
energy density since the continuum excitations, i.e. the string degrees
of freedom, are not included (full dots  in the right panel of
Fig.~\ref{Fig7}), whereas the box calculation with strings gives the
limiting temperature (full squares in Fig.~\ref{Fig7}).  Both curves in
Fig.~\ref{Fig7} have been calculated for density $\rho_0$.

To reproduce qualitatively our box result within the SM we have to include
continuum excitations in the statistical model, i.e.
a Hagedorn mass spectrum for strings \cite{Hagedorn} (for details
see \cite{ThermoNPA}).
For the 'Hagedorn' temperature $T_H$ we use the temperature $T_s$ as
obtained from the box calculations, i.e. $T_H=T_s\simeq 150$~MeV.
As seen in right panel of Fig.~\ref{Fig7} we achieve agreement of the
extended SM and our box calculations.
However, we point out that the limiting
temperature $T_s$ from our string model involves somewhat different
physics assumptions than the Hagedorn model at temperature $T_H$. $T_s$
should not really be identified with the 'Hagedorn' temperature $T_H$,
though close similarities exist. In the Hagedorn picture and for
temperatures close to $T_H$ the abundance of `normal' hadrons or known
resonances stays constant with increasing energy density whereas the
number and energy density of the (hypothetical) bootstrap excitations
diverges for $T\rightarrow T_H$. The Hagedorn model thus assumes
`particles' of mass $m\to\infty$ to be populated for $T\to T_H$, that
dynamically can be formed in collisions of high mass hadrons for
$t\to\infty$.  In contrast, our string model does not include energetic
string-string interactions that might produce more massive strings.

\section{SUMMARY}

In this contribution we have performed a systematic study of equilibration
phenomena and equilibrium properties of 'infinite' hadronic matter as
well as of relativistic nucleus-nucleus collisions using a BUU
transport model that contains resonance and string degrees-of-freedom.
The 'infinite' hadron matter is modelled  by initializing the system at
fixed baryon density, strange density and energy density by confining
it in a cubic box with periodic boundary conditions \cite{ThermoNPA}.

We have shown that the equilibration times $\tau_{eq}$ for different
particles depend on baryon density and energy density. The time
$\tau_{eq}$ for non-strange particles is much shorter than for
particles including strangeness; for kaons and antikaons the
equilibration time is found to be larger than $\simeq$ 40 fm/c for all
baryon and energy densities considered. The overall abundance of the
dominant strange particles (kaons and $\Lambda$'s) being produced and
obtained within the BUU cascade model for heavy-ion collisions can
therefore not be described by assuming a perfect chemical equilibrium
as strangeness is typically still undersaturated to a quite large
extent. We mention that transport model calculations like ours can
describe the yield and spectra of the produced nonstrange hadrons as
well as $K^+, K^-, \Lambda$ yields quite well at SPS energies
\cite{Cass99,Geiss}.  On the other hand, at AGS energies the measured
$K^+/\pi^+$ ratio in central Au~+~Au collisions is underestimated by
about 30\% \cite{CassQM}.  However, we have to point out that the more
exotic strange particles (like the measured antihyperon yields of
Ref.~\cite{WA97}) can by far not be explained within such standard
hadronic multiple channel reactions.  These hadronic data possibly
point towards new physics.

We have, furthermore, shown that thermal equilibrium is established
quickly, within about 5 fm/c at SIS energies and samewhat larger times
at high energies.  The inclusion of continuum excitations, i.e. hadron
'strings', leads to a limiting temperature of $T_s \simeq 150$~MeV in
our transport approach which practically does not depend on the baryon
density and energy.  We have compared our results with the
statistical model (SM), which contains the same degrees of freedom and
the same spectral functions of particles as our transport model. We
found that the limiting temperature behaviour can be reproduced in the
statistical model only after including continuum excitations of the
Hagedorn type, otherwise the fireball temperature extracted from the
particle abundances and spectra is overestimated substantially.

Close to the critical temperature $T_s$, the hadronic energy densities
can increase to a couple of GeV/fm$^3$. From lattice QCD calculations
one expects that a phase transition to a potentially deconfined QGP
state should occur. Referring to the limiting temperature $T_s\approx
150 $ MeV obtained, a QGP should be revealed and clearly distinguished from
a hadronic state of matter if one can unambiguously prove the
existence of an equilibrated and thermal phase of strongly interacting
matter with temperatures exceeding, e.g., 200 MeV.  The best candidates
are electromagnetic probes, either direct photons or dileptons. On the
other hand these are also `contaminated' by hadronic background and/or
preequilibrium physics. So far no thermal electromagnetic source with
temperatures larger or equal than 200 MeV has been clearly identified.
This might happen at RHIC energies in central Au~+~Au collisions which
are expected to be studied soon.


\end{document}